\documentclass[pdflatex,sn-mathphys-num]{sn-jnl}% Math and Physical Sciences Numbered Reference Style 
\usepackage{graphicx}%
\usepackage{multirow}%
\usepackage{amsmath,amssymb,amsfonts}%
\usepackage{amsthm}%
\usepackage{mathrsfs}%
\usepackage[title]{appendix}%
\usepackage{xcolor}%
\usepackage{textcomp}%
\usepackage{manyfoot}%
\usepackage{booktabs}%
\usepackage{algorithm}%
\usepackage{algorithmicx}%
\usepackage{algpseudocode}%
\usepackage{listings}%
\theoremstyle{thmstyleone}%
\theoremstyle{thmstyletwo}%

\theoremstyle{thmstylethree}%

\begin{document}

\title[Article Title]{Securing the Diagnosis of Medical Imaging: An In-depth
Analysis of AI-Resistant Attacks}


\author*[1]{\fnm{ Md Abdullah } \sur{Al Nasim}}\email{ nasim.abdullah@ieee.org }
\equalcont{These authors contributed equally to this work.}

\author[2]{\fnm{Parag } \sur{Biswas}}\email{text2parag@gmail.com}
\equalcont{These authors contributed equally to this work.}

\author[1]{\fnm{Abdur} \sur{Rashid}}\email{A.Rashid.653@westcliff.edu}

\author[3]{\fnm{Kishor Datta} \sur{Gupta}}\email{kgupta@cau.edu}
\equalcont{These authors contributed equally to this work.}

\author[3]{\fnm{ Roy} \sur{George}}\email{george@cau.edu}
\equalcont{These authors contributed equally to this work.}

\author[4,]{\fnm{Sovon} \sur{Chakraborty}}\email{sovon.chakraborty@ulab.edu.bd}

\author[1]{\fnm{ Khalil} \sur{Shujaee}}\email{ kshujaee@cau.edu}
\equalcont{These authors contributed equally to this work.}

\affil*[1]{\orgdiv{Research and Development Department}, \orgname{Pioneer Alpha}, \orgaddress{\city{Dhaka},\country{Bangladesh}}}

\affil[2]{\orgdiv{Department of MSEM}, \orgname{Westcliff University}, \orgaddress{\city{California},  \country{United States of America}}}

\affil[3]{\orgdiv{Department of Computer and Information Science}, \orgname{ Clark Atlanta University,}, \orgaddress{\street{Street}, \city{Georgia},  \country{United States of America}}}

\affil[4]{\orgdiv{Department of Computer Science and Engineering}, \orgname{University of Liberal Arts Bangladesh,}, \orgaddress{\street{Street}, \city{Dhaka},  \country{Bangladesh}}}


\abstract{Machine learning (ML) is a rapidly developing area of medicine that uses significant resources to apply computer science and statistics to medical issues. ML's proponents laud its capacity to handle vast, complicated, and erratic medical data. It is well known that adversaries can force misclassification by constructing inputs maliciously into machine learning classifiers. In the field of computer vision applications, such adversarial examples have been thoroughly researched. Healthcare systems are thought to be highly difficult because of the security and life-or-death considerations they include, and performance accuracy is very important. Recent arguments have suggested that adversarial attacks could be made against medical image analysis (MedIA) technologies because of the accompanying technology infrastructure and powerful financial incentives. Since the diagnosis will be the basis for important decisions, assessing how strong medical DNN tasks are against adversarial attacks is essential. Simple adversarial attacks have been taken into account in several earlier studies. However, DNNs are susceptible to more risky and realistic attacks. The present paper covers recent proposed adversarial attack strategies against DNNs for medical imaging as well as countermeasures. In this study, we review current techniques for adversarial imaging attacks, detections. It also encompasses various facets of these techniques and offers suggestions for the robustness of neural networks to be improved in the future.}




\maketitle

\section{Introduction}\label{sec1}
In recent years, machine learning has been used as a transformative tool, particularly in medical image analysis. Particularly the processing ability of vast and complex medical data showed potential diagnostic accuracy and gave better outcomes. However, though there are advancements in machine learning security, there are several situations of adversarial attacks. These attacks, where inputs are maliciously altered to deceive the model, have been widely studied in computer vision but are less explored in the domain of medical imaging. This paper aims to explore recent adversarial attack strategies targeting DNNs in medical imaging, review detection techniques, and propose potential countermeasures to enhance the resilience and security of ML models in this critical field. We will go through the necessary sections to provide a critical review of these paper. The contributions of this research paper can be summarized below:

\begin{enumerate}
    \item The research provides an in-depth overview of various adversarial attack strategies targeting deep neural networks (DNNs) in medical image analysis to identify vulnerabilities.
    \item It examines existing DNN models used in medical image analysis and their resilience to adversarial threats.
   \item The study evaluates the effectiveness of these DNN models against adversarial attacks.
    \item Recommendations for enhancing security systems in medical image analysis are also provided.
\end{enumerate}

\subsection{Adversarial Attacks ML Applications}
Adversarial attacks are a group of methods for altering machine learning models by adding expertly designed, frequently undetectable changes to input data. These alterations are intended to trick the model, causing it to misclassify data or forecast outcomes \cite{xu2020adversarial}. Adversarial assaults can endanger the dependability and security of machine learning systems, especially in crucial applications like autonomous vehicles, medical diagnosis, and cybersecurity. A number of adversarial attack types exist, but the two most popular subtypes are "White-box attacks" and "Black-box attacks." An continuing area of research in the realm of machine learning security is the analysis of adversarial assaults and defense techniques. Researchers are always exploring novel attack tactics to find possible weaknesses in AI systems while also attempting to create robust models that are less vulnerable to adversarial perturbations. Computer vision, natural language processing, automated driving, and other industries have all benefited greatly from the application of artificial intelligence technology in recent years \cite{qiu2019review}. The uses of artificial intelligence (AI) technology in important security sectors are, however, constrained by the adversarial assaults that AI systems are susceptible to. Therefore, enhancing the resilience of AI systems against adversarial attacks has become more and more crucial to the advancement of AI \cite{qiu2019review}. Artificial intelligence technology applications have recently advanced quickly in many different industries. Artificial intelligence technologies have been used in fields such as image classification, object detection, voice control, machine translation, and more advanced ones like drug composition analysis \cite{ma2015deep}, brain circuit reconstruction \cite{helmstaedter2013connectomic}, particle accelerator data analysis \cite{ciodaro2012online}, \cite{adam2015higgs}, and DNA mutation impact analysis \cite{xiong2015human} due to their high performance, availability, and intelligence. Since Szegedy et al.'s \cite{szegedy2013intriguing} hypothesis that neural networks are susceptible to adversarial attacks, research on artificial intelligence adversarial technologies has steadily become a hotspot, with new adversarial attack techniques and mitigation strategies being developed on a regular basis. The term "adversarial attacks at the training stage" refers to actions taken by the adversaries to alter the training dataset, input characteristics, or data labels during the training phase of the target model. By changing or removing training data, Barreno et al.'s \cite{barreno2006can} alteration of the training dataset falls under the category of training dataset modification.   Adversarial attacks can be categorized into white-box attacks and black-box attacks during the testing stage \cite{qiu2019review}. In white-box scenarios, the attackers have access to the target model's parameters, methods, and structure. Adversaries can use this knowledge to create adversarial samples for attacks.

\subsection{The Adversary's Objective: Evasion attack versus poisoning assault}

The term "poisoning attacks" refers to the assaulting techniques that enable an attacker to insert/modify a number of fictitious samples into a DNN algorithm's training database.
The trained classifier may perform poorly as a result of these bogus data. They may have poor accuracy \cite{biggio2012poisoning} or make incorrect predictions on some test samples \cite{zugner2018adversarial}.
The classifiers used in evasion attacks are fixed and often work well on safe testing samples. The adversaries are unable to alter the classifier's settings or parameters, but they do create some fictitious samples that the classifier is unable to distinguish.

\subsection{Deep Neural Networks: Adversarial Attacks}
Deep neural networks (DNNs) have gained popularity for medical image processing applications like cancer diagnosis and lesion detection \cite{ma2021understanding}. However, a recent study shows that adversarial examples/attacks with tiny, barely noticeable changes can weaken medical deep learning systems. The use of these devices in clinical settings is now accompanied by safety worries \cite{ma2021understanding}.  Deep neural networks (DNN) are increasingly well-liked and effective at a variety of machine learning applications. They have been used with surprising effectiveness in a variety of recognition issues in the fields of pictures, graphs, text, and voice \cite{xu2020adversarial}. They are able to identify things in images with accuracy that is almost human \cite{krizhevsky2012imagenet}, \cite{he2016deep}. Additionally, they are employed in speech recognition \cite{hinton2012deep}, natural language processing \cite{hochreiter1997long}, and gaming \cite{silver2016mastering}. 

\begin{figure}[H]
\includegraphics[width=12.5 cm]{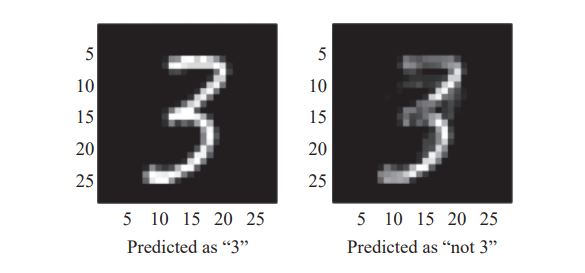}
\caption{The attack of Biggio's SVM classifier for letter recognition. \cite{xu2020adversarial}
\label{fig1}}
\end{figure}   
\unskip

On the MNIST data set, Biggio et al.\cite{biggio2013evasion} produce adversarial instances first, focusing on traditional machine learning classifiers like SVMs and 3-layer fully-connected neural networks which is shown in Figure \ref{fig1}. In order to trick the classifier, it optimizes the discriminant function. As an illustration, consider a linear SVM classifier on the MNIST dataset. 

\subsection{Examining examples of opposition in the real world}

The research \cite{eykholt2018robust} put decals to traffic signs that pose a serious threat to autonomous vehicles' sign recognition technology. These hostile objects can directly interfere with many real-world DNN applications, such as face recognition, driverless vehicles, etc., making them more harmful to deep learning models. By determining if the created adversarial images (FGSM, BIM) are "robust" under natural change (such as changing viewpoint, illumination, etc.), the authors of the work \cite{kurakin2016adversarial} investigate the viability of creating tangible adversarial objects. Robust in this context means that even after transformation, the produced pictures are still antagonistic. The experimental findings show that a significant fraction of these adversarial examples, particularly those produced by FGSM, continue to be adversarial to the classifier following transformation. The findings point to the possibility of actual hostile objects that can trick the sensor in various settings.

\subsection{Medical Image Under Artificial Intelligence Attack}

Researchers have access to strong models of developing science and technology thanks to deep learning. As they are highly good at learning useful features, convolutional neural networks (CNNs) are the most significant class of DL models for image processing and analysis.  Medical image analysis is the processing of the human body using various picture modalities for diagnostic, therapeutic, and health monitoring purposes \cite{apostolidis2021survey}. The development of deep neural networks in the field of computer vision addresses issues that were not well-solved by traditional image processing methods. Beinfeld et al. \cite{beinfeld2005diagnostic} asserted that \$385 spent on medical imaging results in a savings of almost \$3000. MRI, CT scans, ultrasound (US), and X-rays are the most frequently used image modalities. Figure 1 illustrates how diagnosis outcomes can be arbitrarily changed by adversarial attacks across three medical picture datasets: Fundoscopy \cite{graham2015kaggle}, Chest X-Ray \cite{wang2017hospital}, and Dermoscopy \cite{jones2019dermoscopy}. 

\begin{figure}[H]
\includegraphics[width=13.5 cm]{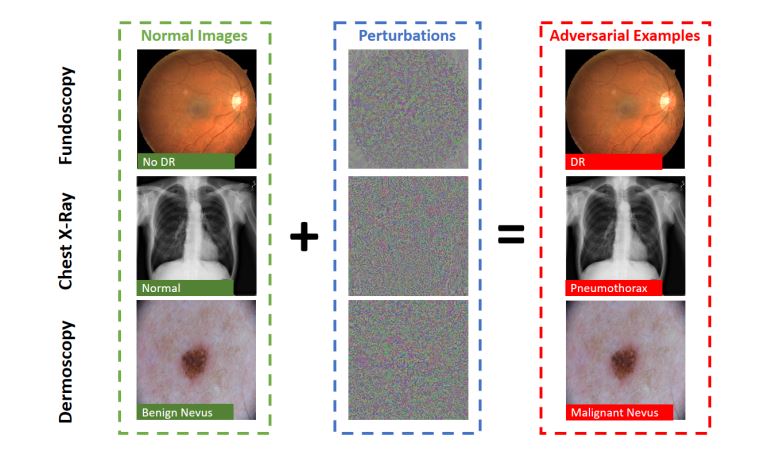}
\caption{Examples of adversarial approaches designed by the Projected Gradient Descent (PGD) to deceive DNNs trained on medical image datasets include dermoscopy \cite{jones2019dermoscopy} (third row), chest x-ray \cite{wang2017hospital}, and fundoscopy \cite{graham2015kaggle} (first row, DR=diabetic retinopathy). Normal images on the left, adversarial perturbations in the middle, and adversarial images on the right. The anticipated class is indicated by the left-bottom tag, and green or red denotes accurate or inaccurate predictions \cite{ma2021understanding}. \label{fig2}}
\end{figure}   
\unskip

Hu et al. \cite{xu2020adversarial} assert that in order to develop more reliable models, it is crucial to investigate the reasons why adversarial cases emerge and to comprehend deep learning models better. Depending on the knowledge of the enemy, attacks can be categorized into three kinds.   Based on the gradients of the classification loss with respect to the input, the saliency (or attention) map of an input image indicates the regions that significantly alter the model's output \cite{ma2021understanding}. We can see that some medical photos have highly concentrated regions that are noticeably larger. This could mean that the DNN model is occasionally drawn away by the rich biological textures present in medical images, causing it to focus more on details unrelated to the diagnosis. Small modifications in these areas of high focus can have a big impact on the model's output. 

Due to several factors, including High-Dimensional Data, High-Dimensional Data, Transferability of Adversarial Examples, Complex Decision Boundaries, Black-Box Attacks, Safety-Critical Applications, and Safety-Critical Applications, Medical Image Deep Neural Network (DNN) models can be relatively easier to attack when compared to some other domains. Despite these difficulties, experts in the field are actively attempting to create more reliable and safe DNN models for medical images. To make these models more resistant to hostile attacks, strategies like adversarial training, input preprocessing, and defensive distillation are being investigated. To further enhance the security and dependability of AI systems in healthcare applications, consistent evaluation frameworks for adversarial robustness in medical imaging must be developed.

An attack method on ultrasound (US) imaging for fatty liver was put forth by Byra et al. \cite{byra2020adversarial}. Radio-frequency signals are used to rebuild US pictures, and the reconstruction technique was subjected to a zeroth-order optimization attack \cite{chen2017zoo}. The InceptionResNetV2 model was used in the studies, and the assault resulted in a 48\% loss in model accuracy. Adaptive segmentation mask attack (ASMA), a specific attack for medical picture segmentation, was suggested by Ozbulak et al. \cite{ozbulak2019impact}. The suggested attack offers significant intersection-over-union (IoU) degradation and produces nearly undetectable samples for most portions. Because the U-Net model is one of the most well-known models for medical picture segmentation, they employed it in the trials. Datasets for ISIC skin lesion segmentation \cite{codella2018skin} and glaucoma optic disk segmentation \cite{pena2015estimation} were employed.  A technique for creating hostile instances to undermine medical picture segmentation was put out by Chen et al. \cite{chen2019intelligent}. In order to simulate anatomical and intensity fluctuations, geometrical deformations are used to create the adversarial examples. By attempting to partition organs from abdominal CT scans using a U-Net model, they tested the effectiveness of these examples. In terms of the Dice score metric, they successfully reduced it significantly across all organs. The kidneys and pancreas, however, require a higher level of disturbance and are more difficult to assault than the liver and spleen.

\begin{figure}[H]
\includegraphics[width=12.5 cm]{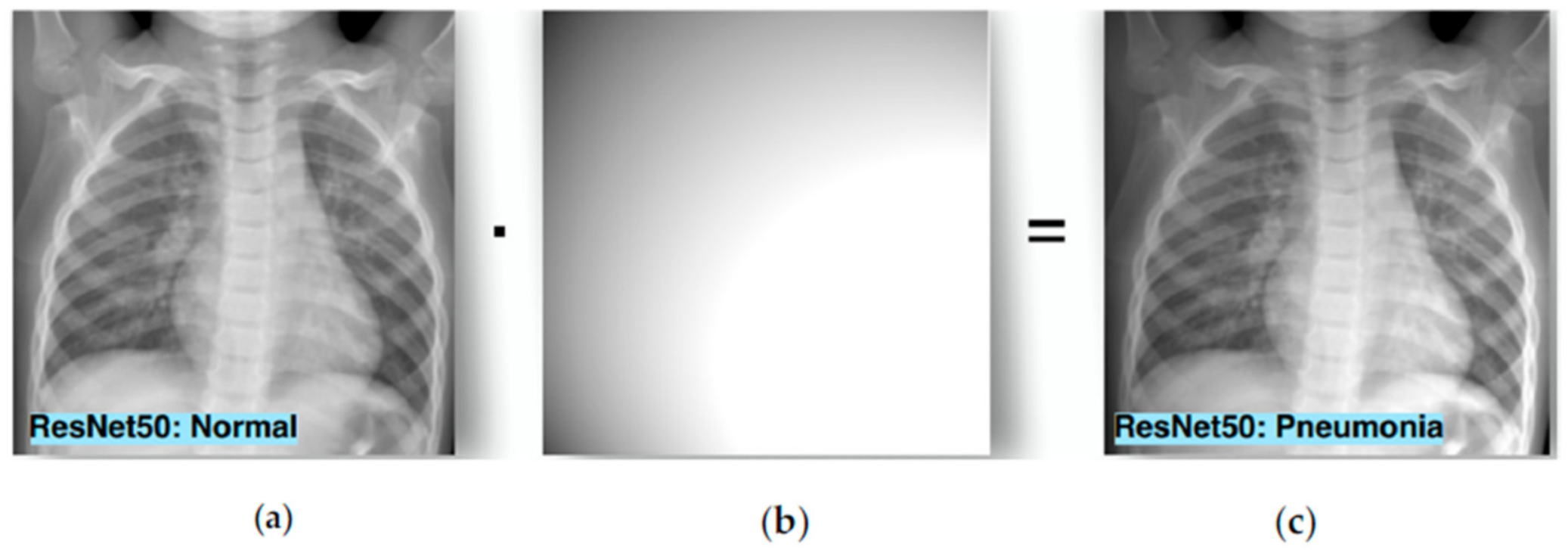}
\caption{Clean image, bias field noise, and diagnosis following application of bias field noise are shown in (a), (b), and (c), respectively \cite{tian2021bias}. 
\label{fig3}}
\end{figure}   
\unskip

As seen in Figure \ref{fig3}, Tian et al. \cite{tian2021bias} looked at the phenomenon of bias field, which can result from improperly acquiring a medical image and compromise a DNN's effectiveness. The authors developed an adversarial-smooth bias field attack to deceive a model after being inspired by adversarial assaults. The chest X-ray dataset used for their research was tuned using the ResNet50, MobileNet, and DenseNet121 models. They looked at this attack's transferability and white-box attacks. Comparing the suggested attack to other cutting-edge white-box attacks, it exhibited a greater attack accuracy on transferability.

This review's major organization is as follows: We offer some significant adversarial notions on medical images in Section 2. Additionally, it provides a detailed overview of earlier research. We explore a few strategies in Section 3 with regard to the picture classification scenario. We utilize Section 4 to quickly summarize some findings from earlier research that tries to explain the phenomena of antagonistic examples on medical data. The overview is concluded in Section 5.

\section{ Background Literature Review}

Adversarial examples are data inputs that are specifically designed to make a machine learning model fall short. The term "adversarial inputs" was first used informally in 2004, when academics looked into the strategies spammers used to get around spam filters \cite{dalvi2004adversarial}. Typically, threatening examples are created by intentionally altering genuine data, such as spam advertising messages, in order to deceive the algorithm that will process it. Such modifications can be made to text data, such as spam, by adding innocent text or by changing phrases that are frequently used in malicious communications to synonyms. Researchers have worked to create algorithms that are resistant to adversarial attacks, for example by exposing algorithms to hostile examples during training or utilizing cunning data processing to reduce the possibility of manipulation. Early efforts in this direction are encouraging, and we are hopeful that the pursuit of completely robust machine-learning models will spur the creation of algorithms that learn to make judgments for consistent justifications \cite{zhang2023self}. 

Following payer approval, medical claims codes determine the amount paid for a patient visit. Payers often use automated fraud detectors, which are increasingly driven by machine learning, to assess these claims. Health care providers have historically shaped payers' records of patient visits (and the related codes) to affect their decisions (the algorithmic outputs) \cite{zhang2023self}.  Medical fraud, a \$250 billion market, is at the extremity of this tactical tailoring of a patient presentation. Although some practitioners might blatantly fabricate medical claims, patient data falsification frequently takes much more covert forms. Intentional upcoding, for instance, is the practice of routinely submitting billing codes for services that are similar to but more expensive than those that were actually rendered. 

\subsection{The components of an adversarial attack}

The goal of the study \cite{puttagunta2023adversarial} is to present a comprehensive overview of the various adversarial assault strategies and defense techniques. First, we talk about the theoretical foundations, procedures, and applications of adversarial attack tactics. The research on defense strategies covering the large field's boundary is then briefly discussed. Figure \ref{fig5} shows the year-by-year publishing of a few chosen articles from 2012 to 2021. In-depth taxonomies and analyses of harmful attacks and defenses against the full DL pipeline are the goals of this research. The numerous existing attack and defense strategies were categorized in this context, with a focus on those that have only been developed in the last two years, and the medical deep learning systems sensitive to adversarial attacks were examined. 

Numerous defensive strategies have been suggested as a defense against the threat. Adversarial training, which enlarges the training dataset with adversarial images to improve the resilience of the trained Convolutional Neural Network (CNN) model, is a common technique in the natural imaging domain. This tactic is not ideal for medical imaging datasets, though, as the introduction of a large number of different hostile images into the training set can seriously reduce the classification accuracy.  In article \cite{li2020robust}, authors suggest a reliable detection method for malicious images that can successfully fend off attacks on deep learning-based medical image categorization systems. In the study \cite{ghaffari2022adversarial}, scientists looked at how resilient CNNs in computational pathology were to various attacks and compared the findings to how resilient ViTs were. Authors also developed strong neural network models and assessed how well they defended against white- and black-box attacks. The attack structures for both models were examined, and the authors looked into why they performed as they did. The findings were verified by the authors using two clinically pertinent classification tasks on separate patient groups. Prior to the publication of Ma et al.'s study \cite{ma2021understanding}, medical image AAs were ambiguous and adversarial machine learning analysis was limited to natural image analysis. Medical images, in contrast to natural images, may include domain-specific elements. Medical deep-learning systems may be impacted by AAs. AAs have the discretion to alter diagnoses and results. Ma et al. emphasized the significance of the healthcare system's extensive resource base. It invariably raises dangers where potential attackers might try to take advantage of the healthcare system. 

\section{Materials and Methods}
This study examines various strategies to strengthen the usability of medical image analysis models against adversarial attacks. These strategies include techniques such as input denoising, adversarial training, adversary detection, image-level pre-processing, feature enhancement, and knowledge distillation. Each method tries to improve the aspects of models, which is necessary to the potential risks these attacks pose to clinical applications. Additionally, the research presents a new classification system for adversarial defense in medical imaging, categorized by application type, covering white-box, semi-white-box, and black-box attacks. By addressing the vulnerabilities in computerized diagnostic systems, this work contributes to the development of reliable and accurate healthcare technologies, ensuring consistent performance even under adversarial conditions.
\subsection{Medical Adversarial Defensive Strategies}
Defense models such as input denoising, input gradients regularization, and adversarial training have been created. The most recent attacks can typically completely or partially escape these countermeasures. In this case, a thorough investigation of adversarial attack and defense techniques in the field of medical image analysis is presented, including a family of techniques with a novel taxonomy based on the application situation \cite{dong2023adversarial}. A unified theoretical framework was built by the author for various adversarial attack and defense strategies in medical imaging applications.
The white-box scenario is the main focus of most efforts on adversarial attack for medical image processing. These studies, in particular, with complete understanding of medical DNNs, focus on the adversarial vulnerability of computer-aided diagnosis models in diverse medical imaging applications. When executing adversary creation, the attacker might use the target diagnosis DNN as a locally deployed model.Additional academics have looked into the weaknesses of additional medical imaging tasks in addition to white-box adversarial attacks for medical classification tasks. These publications mostly focus on medical segmentation \cite{chen2019intelligent} \cite{ozbulak2019impact}. 
For natural photos, semi-white-box (Gray-box) attacks have been extensively researched \cite{xiao2018generating}. However, there are only a few academic publications \cite{rahman2020adversarial}, \cite{wang2022feature} that address this attack scenario for medical image processing. The semi-white-box adversarial approach typically consists of two stages: 1) To create adversarial instances against the target DNN model, the attacker trains a generative model. The attacker has complete access to the target model during training, including backward propagation gradients. 2) Instead of needing to know anything about the target model, as would be the case in a completely black-box scenario, the adversary generator can directly obtain adversarial examples against the target model with the input of authentic photos during the application stage.
Currently used white-box adversarial attacks mostly call for numerous target model backward gradients. In other words, the attacker generates equivalent adversarial samples by treating the target DNN as the locally deployed model. However, because it requires a comprehensive understanding of the DNN model to attack, the white-box option can be unreliable in a real-world scenario. Comparatively, the general black-box scenario may be a better environment for simulating real-world adversarial attacks. There have been numerous proposals to investigate black-box assaults for natural images. 
The solutions outlined above either require numerous queries to the black-box model or depend on having complete understanding of the desired diagnosis model. However, in the majority of real-world scenarios, the attacker might not be able to directly access the target diagnosis model. The restricted black-box (no-box) configuration in particular might effectively represent the toughest (worst) situation for the real-world adversarial assault, even if it doesn't call for querying the target blackbox DNN. The transferability \cite{liu2016delving} of adversarial cases across different DNN models is the fundamental determinant of the no-box attack. For instance, a no-box attacker can create adversarial images based on a surrogate model that has been deployed locally and that can be transferred straight to the target medical diagnosis systems. Restricted black-box adversarial attacks are more covertly dangerous for natural vision tasks, according to a well-known study. 
Given the significant dangers to the healthcare sector, a number of defense strategies have been put out to fend off medical hostile attacks. Numerous approaches have been put out to provide reliable deep learning-based systems for natural photos in response to the catastrophic failures brought on by adversarial instances \cite{liu2016delving}. Developing accurate computer-aided diagnosis models for clinical applications also helps millions of people receive trustworthy healthcare services. Investigating adversarially robust models in the context of medical image analysis is therefore important. 

In \cite{dong2023adversarial}, authors summarize the works on adversarial defense in the context of medical imagine analysis. In the context of medical pictures, various attack methods also serve as robustness evaluation criteria for adversarial defense, in addition to a considerable number of adversarial attack methods stated here. 

\subsection{Adversarial Training}
Most medical adversarial defense techniques focus on using adversarial training to create reliable diagnoses systems. A significant fraction of the works among them go beyond current techniques for natural image adversarial training to tasks relating to medical classification \cite{xu2021towards}, \cite{paul2020mitigating}, Vatian et al. \cite{vatian2019impact} looked into opposing examples for medical imaging and tried a number of defense strategies to oppose these nefarious representations.

\subsection{Adversarial Detection}
Adversary detection seeks to identify adversary cases from input examples during the application stage, as opposed to developing robustness during the training stage of computer-aided diagnosis models. In the field of biological image analysis, a variety of adversarial detection techniques have been put forth to prevent further misdiagnosis brought on by adversarial examples \cite{li2020robust}. In particular, it is possible to think about medical adversarial detection as an anomaly detection issue that can be resolved by combining explainability approaches \cite{watson2021attack}.

\subsection{Image-level Pre-processing}
A clean image and the related adversarial perturbation make up an adversarial image in general. Meanwhile, it has been shown that DNNs can perform well on clean images while still being vulnerable to adversarial examples \cite{szegedy2013intriguing}. Denoising the adversarial example to remove the perturbation component can therefore help make the subsequent network diagnostic easier. Image-level pre-processing can be useful and secure in the context of biomedical image analysis because it does not require re-training or modifying medical models. 

\subsection{Enhancement of Features}
The discrepancy between the robustness of human and machine vision has been shown to be caused by adversarial examples being associated with non-robust features (extracted from specific patterns in the data distribution) \cite{ilyas2019adversarial}. Therefore, improving the feature representation is absolutely required for robust inference systems. In this study, we define the feature augmentation as the alteration of architectures or mapping functions. The robustness of medical classification models has been increased by the development of numerous feature improvement techniques \cite{han2021advancing}. 

\subsection{Distillation of Knowledge}
Knowledge distillation is a useful method for moving learnt information from a complicated (teacher) model to a simple (student) model in the machine learning field. As a result, self-distillation refers particularly to the scenario in which the network architectures for the teacher and student models are the same. Additionally, adversarial knowledge distillation, which transfers the adversarial resilience from the heavy teacher model to a light student model, has been extensively investigated for the natural imaging domain \cite{goldblum2020adversarially}.

\section{Result and Discussion}

\begin{figure}[H]
\includegraphics[width=13.5 cm]{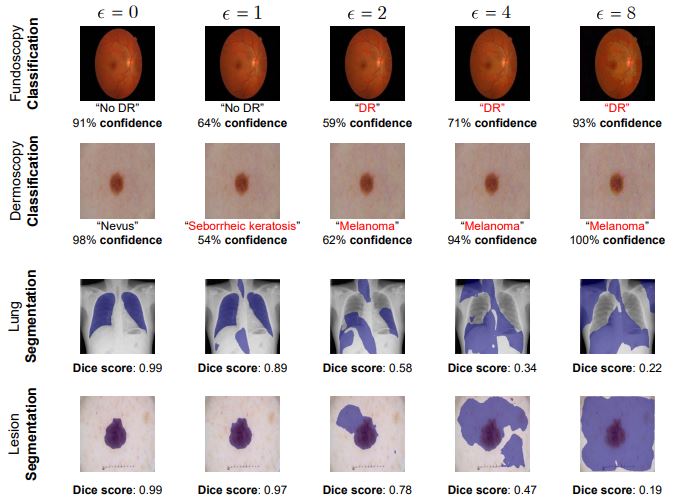}
\caption{Medical adversarial examples with predictions for a range of perturbation sizes. For visualization, the created segmentation masks are placed on top of the source photos \cite{dong2023adversarial}.\label{fig6}}
\end{figure}

Four publicly available benchmark datasets are used in this study \cite{dong2023adversarial} to explore adversarial attack and defense for medical image processing (shown in Figure \ref{fig6}). 2) Messidor1 dataset of 1,200 eye fundus color numerical pictures for detecting diabetic retinopathy of four classes according to retinopathy grade; International Skin Imaging Collaboration (ISIC) dataset of 2,750 dermoscopic images of three different categories for skin lesion classification and segmentation. 3) The ChestX-ray 14 dataset consists of 112,120 frontalview X-ray images from 14 thorax diseases. 4) The COVID-19 database, which includes 21,165 chest X-ray images with segmentation-capable lung masks.

\unskip

Experiments are used by authors \cite{li2021defending} to show: 1): Without affecting the classification accuracy of clean images, the SSAT module can considerably improve the adversarial robustness of the model. 2) The bulk of successful adversarial examples may be found and excluded by the UAD module. 3) When compared to other existing AI systems, their medical imaging AI solution (UAD + SSAT) minimizes adversarial risk. A publicly available dataset of retinal OCT images is used for the research. A publicly available dataset of retinal OCT images is used for the research.  The most difficult threat used to assess class prediction performance is the "white-box" setting.In all white-box attack conditions, the authors show that SSAT significantly outperforms alternative baselines while maintaining an equivalent or superior performance for clean picture classification. The authors show that the lowest adversarial risk for the new measure presented results from UAD complementing with SSAT.  Regardless of the training techniques employed, it is evident that UAD-based systems consistently have reduced risks compared to those who do not.
\begin{figure}[H]
\includegraphics[width=13.5 cm]{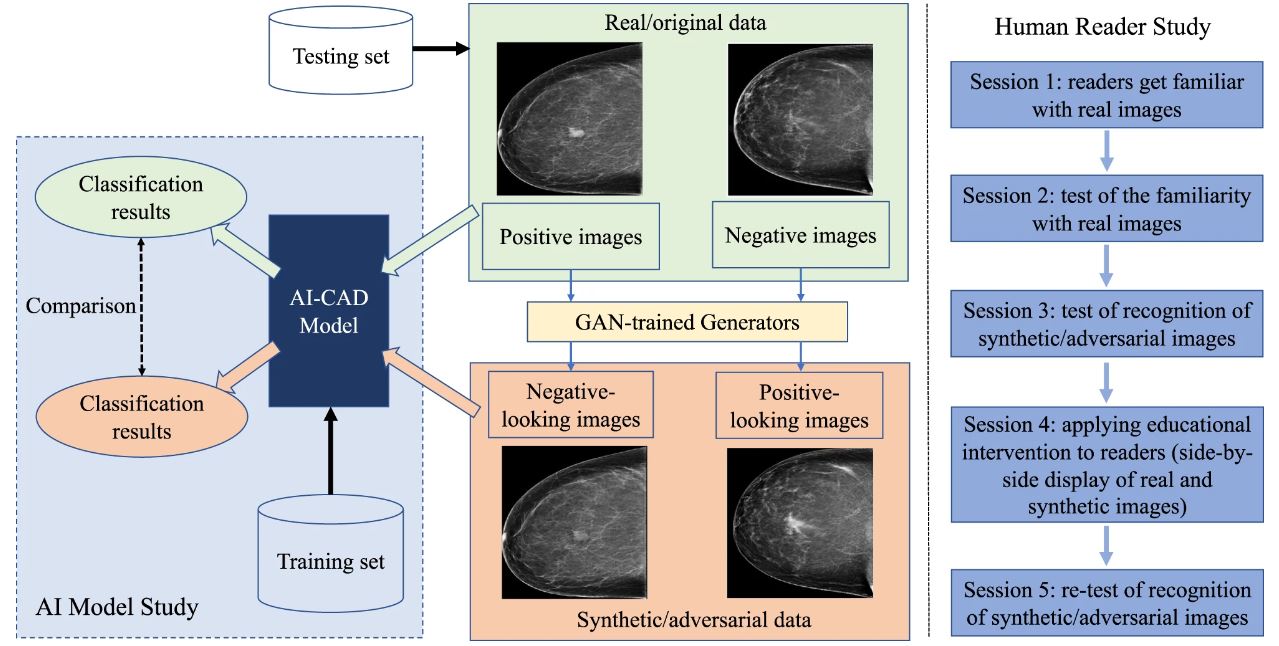}
\caption{An outline of our study's methodology. \cite{li2021defending}\label{fig7}}
\end{figure}   
\unskip
An AI-CAD model, in Figure \ref{fig7},  that intended to alter the diagnosis-sensitive contents of images (by adding or deleting malignant tissue) was first trained, and then it was evaluated on the adversarial images produced by the GAN model. The reader study looked at how well human specialists could visually identify the images created by the GAN.
Authors \cite{zhou2021machine} of a research paper conducted a study to assess an AI-CAD model's responses to adversarial attacks on GAN-generated mammography pictures by introducing malignant tissue into healthy photos and removing cancerous tissue from cancer-affected images. Additionally, they assessed the effectiveness of experienced radiologists in recognizing these types of antagonistic pictures visually both without and after an instructional intervention. The University of Pittsburgh Medical Center provided the authors with 1284 women for their study cohort, and they also collected 4346 mammography pictures for this cohort. In this cohort, 366 patients had breast cancer that was biopsy-proven to be malignant while 918 patients had breast cancer that was evaluated as benign (including benign signs). Following training, researchers evaluated the model's classification accuracy in comparison to both the original genuine test data and its equivalent GAN-generated adversarial counterparts. To generate the adversarial images, we created two GAN-trained U-Net23 models. The GAN generators, which were trained to create fake positive mammography images from the falsely negative mammogram photos and fake negative mammogram images from the falsely positive mammogram images, respectively, updated the mammogram images in the test set with flipped labels. The two distinct resolutions of this image indicate the categorization impacts of the AI-CAD model on the test data. The authors begin by outlining the findings from the high-resolution (1728 1408) photographs. As can be observed, the AI-CAD model obtained an AUC of 0.82 on the test set, which consisted of 364 genuine negative samples and 74 real positive samples. The test set's equivalent adversarial GAN-generated images (with flipped labels) yielded an AUC of 0.94. These two AUC values show that the adversarial collection of photos generally succeeded in deceiving the AI-CAD model. Additionally, when calculating classification accuracy using a threshold of 0.5, 44 out of the 74 real positive images (or 59.5\%) were correctly labeled as positive cases; however, only 42 out of the 44 fake negative images (note that their labels were negative due to the flipping) were categorized as negative cases, indicating that 95.5\% (42 out of 44 cases) of the GAN-generated adversarial samples successfully deceived the system.

\begin{figure}[H]
\includegraphics[width=13.5 cm]{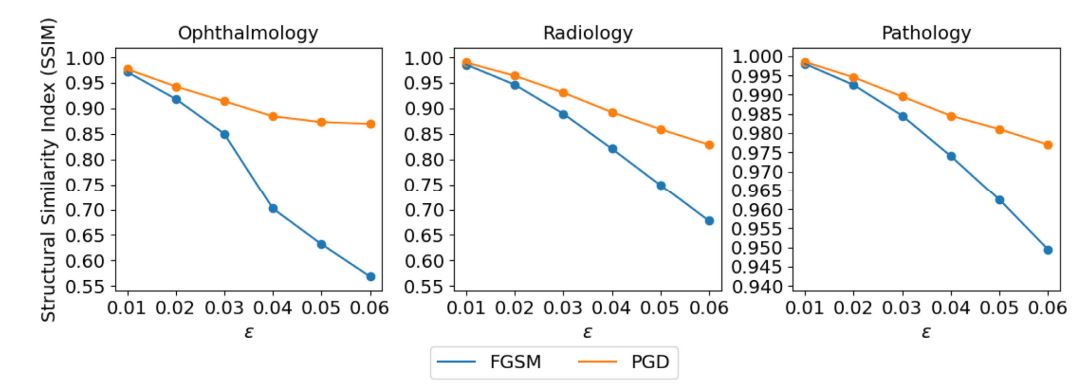}
\caption{Between the original images in the test sets and the adversarial examples produced using FGSM or PGD with varied perturbation degrees, the Mean Structural Similarity Index Measure (SSIM) was calculated. The SSIM values shown by the points are averaged over two model architectures (Inception-v3 and Densenet-121).\cite{bortsova2021adversarial}
\label{fig8}}
\end{figure}   
\unskip

Figure \ref{fig8} displays the mean SSIM values for FGSM and PGD assaults across all photos. The Supplementary Material includes the SSIM data for every model. As can be shown, the effects of the identical disturbance applied to various imaging modalities on human visual perceptibility with the observed SSIM varies. The radiology images most clearly showed adversarial disturbances, with = 0.02 producing an already apparent, but very modest perturbation. At the same perturbation level for the ophthalmology and pathology photos, perturbations were nearly undetectable and became apparent with larger epsilon values. Black-box adversarial assaults on deep learning in medical imaging are studied by the authors \cite{bortsova2021adversarial}. They investigate three medical imaging sectors where deep learning algorithms are vulnerable. In all three datasets, perturbations calculated by FGSM showed lower SSIM than those calculated using PGD. This is a predicted outcome given that PGD optimizes perturbations based on their size and influence on model predictions.We decided to report attacks using for our further investigations utilizing = 0.02, since this was the maximum degree of disturbance that was still for all applications and assault strategies, and it had more transferability than an epsilon of 0.01 in most cases of the applications examined.

\begin{figure}[H]
\includegraphics[width=13.5 cm]{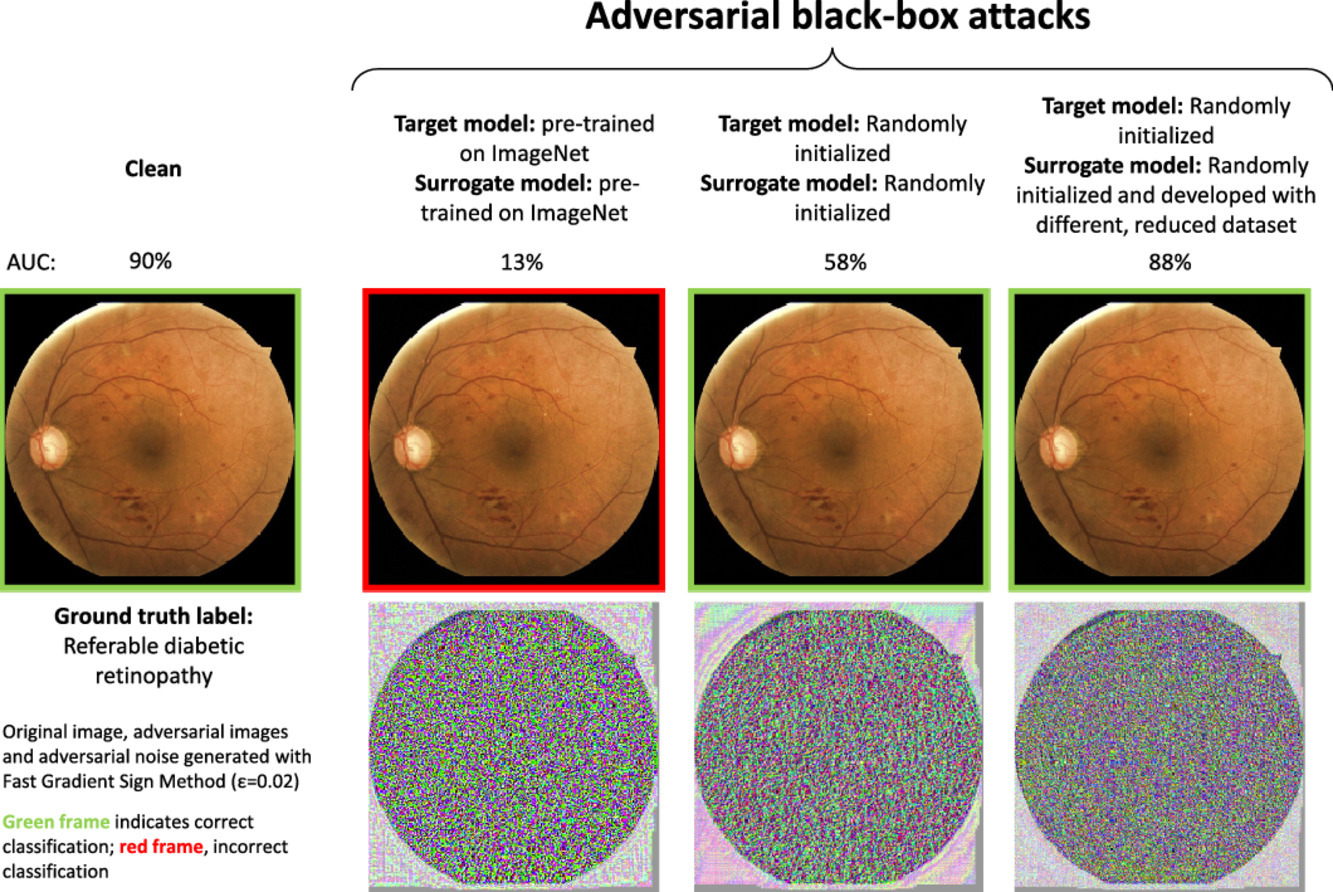}
\caption{FGSM (=0.02) was used to create the original images, adversarial images, and corresponding adversarial noise in a variety of black-box settings, including target and surrogate pre-trained on ImageNet, target and surrogate randomly initialized, target and surrogate randomly initialized plus surrogate developed using a different and reduced dataset (d2/2). The average receiver operating characteristic curve (AUC) for each configuration for both clean and black-box scenarios is shown above. Diabetic retinopathy (DR) is correctly classified as referable or non-referable in a green frame and incorrectly classified in a red frame. The difference between the original and the adversarial image is represented by the adversarial noise \cite{bortsova2021adversarial}. 
\label{fig9}}
\end{figure}   
\unskip

Examples from the ophthalmology dataset are shown in Figure \ref{fig9} to demonstrate attack transferability when the target and surrogate are both pre-trained on ImageNet and when both are initialized randomly.
The instances where the target and surrogate models had the same or different architectures had no effect on any of the aforementioned results. 

\section{Conclusion and Discussion}

The study identifies the impact of adversarial defense strategies in medical image analysis using datasets like Messidor1, ISIC, ChestX-ray14, and COVID-19. The SSAT module significantly improves model robustness against adversarial attacks while maintaining accuracy on clean images. Combined with the UAD module, it efficiently detects and removes adversarial examples, minimizing risks in upcoming applications. Diagnostic imaging AI systems based on computer vision are increasingly being used as models for classifying and segmenting diseases. The scaled-up use of medical imaging AI systems has given rise to serious safety issues because to DNNs' susceptibility to adversarial samples. Recently, a number of ways have been put out to increase the efficacy of medical image defense tactics. Although numerous protection strategies have been put forth, there are still reservations over the use of medical deep learning techniques. This is a result of some limitations in medical imaging, such as a dearth of high-quality picture datasets and labeled data as compared to other high-quality natural image datasets. Adversarial defenses might not have a limited impact. The results of the study highlight the need for greater care in designing DNNs for medical imaging and their practical applications.

\begin{figure}[H]
\includegraphics[width=12.5 cm]{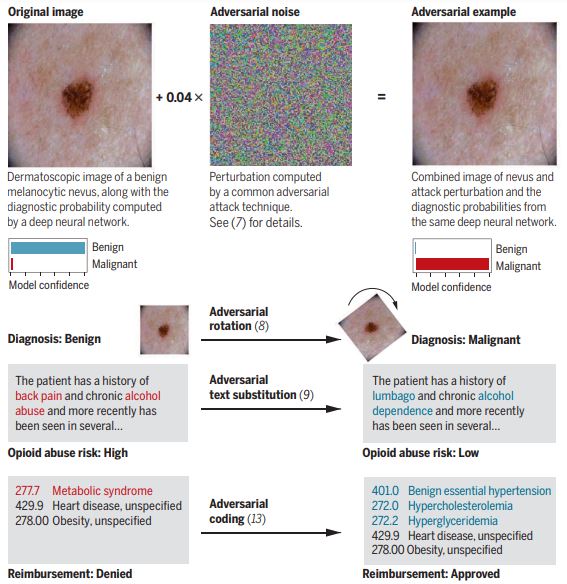}
\caption{A demonstration of how adversarial attacks against different medical AI systems could be carried out without requiring any openly dishonest data manipulation \cite{finlayson2019adversarial}. 
\label{fig4}}
\end{figure}   
\unskip
\bibliography{bibliography}
\end{document}